\documentclass{PoS}

\title{The modeling of the Vela pulsar pulses - from optical to hard gamma-ray energy}

\ShortTitle{The modeling of the Vela pulsar pulse profiles}

\author{\speaker{B. Rudak}\\
        CAMK, Rabia{\' n}ska 8, 87-100 Toru{\' n}, Poland\\
        E-mail: \email{bronek@ncac.torun.pl}}

\author{J. Dyks\\
        CAMK, Rabia{\' n}ska 8, 87-100 Toru{\' n}, Poland\\
         E-mail: \email{jinx@ncac.torun.pl}}

\abstract{The pulsed radiation from PSR B0833-45 (Vela) has a phased-averaged spectral energy distribution
(SED) of an apparently simple structure across a wide energy range, from optical light to hard gamma-rays.
However, the Vela pulses in narrow energy bands reveal astonishing complexity of the directional 
pattern of the radiation. These pulses are, therefore, a unique clue to the underlying radiative processes. 
We present the results of a 3D modeling of the Vela radiation properties within an outer-gap scenario. 
We show how the inverse Compton scattering of photons by primary and secondary particles 
in its magnetic and non-magnetic regimes reproduces qualitatively, and in some cases quantitatively, 
the observed energy-dependent pulses of Vela.}

\FullConference{35th International Cosmic Ray Conference --- ICRC2017\\
		10--20 July, 2017\\
		Bexco, Busan, Korea}

\begin{document}

\section{Introduction}
The Vela pulsar (PSR B0833-45) is one of best studied rotation-powered pulsars in the high-energy radiation domain
due to its proximity. No wonder, there have been many theoretical efforts to explain the mechanisms of its activity,
applying various models.
This paper presents an attempt to model inverse Compton scatterings expected between various sets of relativistic electro-positron pairs operating
in the pulsar magnetosphere and two different radiation fields: 1) non-thermal near-IR to optical emission interpreted as due to synchrotron
emission originating near the accelerating gaps; 2) thermal X-rays from the neutron star surface.
The consequences of those interactions are shown to be present at the optical domain as well as the very-high energy domain (VHE).

\section{The Outer Gap model}
\begin{figure}
\centering
\includegraphics[width=.8\textwidth]{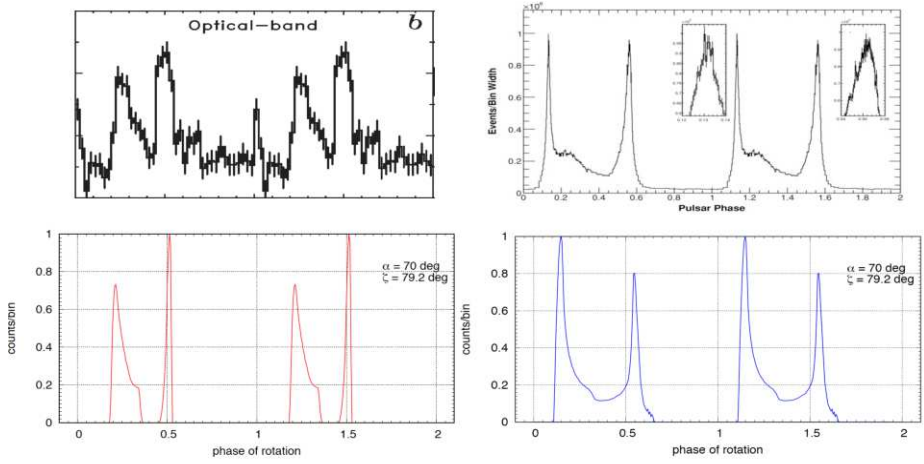}
\caption{Two pronounced peaks in the lightcurves of the Vela pulsar, in optical (\cite{gouiffes} and in gamma-rays above 100 MeV (\cite{abdo})
- upper left panel and upper right panel, respectively - are interpreted as due to synchrotron emission from secondary pairs 
and curvature emission due to primary pairs, respectively. Assuming that the gamma-rays come from the outer gaps of the thickness $0.1$, and the optical light comes
from layers of the same thickness but adjacent to the outer gaps, an auxilary geometrical model was used to determine
the inclination angle $\alpha$ and the viewing angle $\zeta $ of Vela by
reproducing the shapes and rotational phases
of all four peaks. A satisfactory match was found for $\alpha = 70\,$deg and $\zeta = 79\,$deg, and these values were next used in the physical model.}
\label{fig1}
\end{figure}

The physical model of OG used in this work is unavoidably simplified due to the fact that some aspects of  physical conditions in OG
pose  still open questions. We use a standard corotating magnetosphere in the low charge density approximation. The magnetic field 
is taken as a vacuum retarded dipolar solution. The numerical 3D code calculates curvature radiation (CR) from primary charges in the inertial observer frame (IOF)
as described in \cite{d04} and \cite{bai}. The charges (primary pairs) are subject to an accelerating electric field which is assumed
to be constant in the gaps (i.e. no azimutal or radial dependence is considered). The assumed accelerating field $E_{||}$ allowed the primaries to reach 
Lorentz factors of a few $\times 10^7$.  
The model doesn't follow the pair creation due to interaction of  CR photons
with a soft-photon field. Instead, we assume that secondary pairs are formed in a layer which is attached to (placed on top of) the outer gap layers.
These pairs are the source of the soft photon field - synchrotron radiation (SR) which extends across the wide energy range, from hard X-rays to UV and optical up to
midIR, and likely to far IR ($\sim 0.001$ eV). To determine the directional characteristic of the SR radiation which would agree with the observed pulses from the Vela pulsar,
with used the geometrical version of OG (\cite{d04}). The geometrical model assumes a uniform distribution of emissivity in the energy range of interest, with photons emitted tangentially to local magnetic filed lines. Therefore, for
a given location of the layer in terms of the magnetic-line footprints on the polar cap, the pulses of radiation (CR in the case of the gap, or SR in the case
of the secondary-pairs layer) will be a function of two angles: the inclination angle $\alpha$ of the magnetic axis with respect to the spin axis, and the viewing
angle $\zeta$ between the line-of-sight and the spin axis. Taking the thickness of both layers to be the same and equal 0.1 (ie. 10\% of the polar-cap radius)
the best set of $\alpha$ and $\zeta$ was determined as 70 deg and 79 deg, respectively. The resulting 'geometrical' lightcurves of CR (gamma-rays) and SR (optical)
along with the data are presented in Fig.\ref{fig1}.  The optical lightcurve in the geometrical model consists of two peaks located at phases 0.25 and 0.5,
similarly as two major peaks (called P1 and P2, respectively) observed in the lightcurve of Vela (\cite{gouiffes}). Also, the shape and thickness of the model peaks are roughly similar to the observed peaks. The similar situation is for the geometrical and observed lightcurves in the gamma-rays, where two peaks
are located at phases 0.13 and 0.56  (\cite{abdo}). 

Two important ingredients of the physical model are:\\ 
1) The electron-positron pairs of $\gamma \sim 100$  present in the inner part of open magnetic field-lines,
ie. concentrated around the magnetic axis (hereafter 'the inner PC pairs'). The justification for this component it twofold: 
a) the outer gap and its adjacent SR layer occupy about 20\% (in terms of the
polar cap radius) of the  magnetic-field lines at the outer rim of the polar cap; there is no reason to prevent the inner part of polar cap to develope a potential drop across it,
high enough to lead eventually to the creation of electron-positron pairs; b) a narrow radio pulse ("a core" emission) is present in Vela.\\
2) The field of thermal soft X-ray emission from the Vela surface (\cite{manzali}).

Three different ICS processes were considered in an attempt to either predict or to explain the existence of specific features in the Vela's non-thermal
radiation:
\begin{enumerate}
\item primary electrons in the gap interacting with optical-infrared photons originating from the SR layer,
\item inner PC pairs interacting with optical-infrared photons originating from the SR layer,
\item inner PC pairs interacting with thermal X-ray photons from the neutron star surface.
\end{enumerate}

These processes are expected to lead to the formation of, respectively,:
\begin{enumerate}
\item spectral component in the VHE range,
\item core-like pulse (known as P4) detected in optical as well as in UV (\cite{gouiffes}, \cite{uv})  at the phase of the radio core pulse,
\item core-like pulse detected in hard X-rays  (\cite{akh02})  at the phase of the radio core pulse (see also \cite{kuiper} for the compilation of all
lightcurves available).
\end{enumerate}

\section{Results}
\subsection{A  component in the VHE range}
An important role in shaping the resulting directionality of the ICS emission in this case
plays the bending of trajectories of the SR photons in the corotating frame (CF) with respect to the primary electrons moving along magnetic field lines in the gaps.
For high inclination angles this leads to lower rates of photon-electron collisions at at the leading side than at the trailing side (see fig.6 in \cite{d13}).

\begin{figure}
\centering
\includegraphics[width=.7\textwidth]{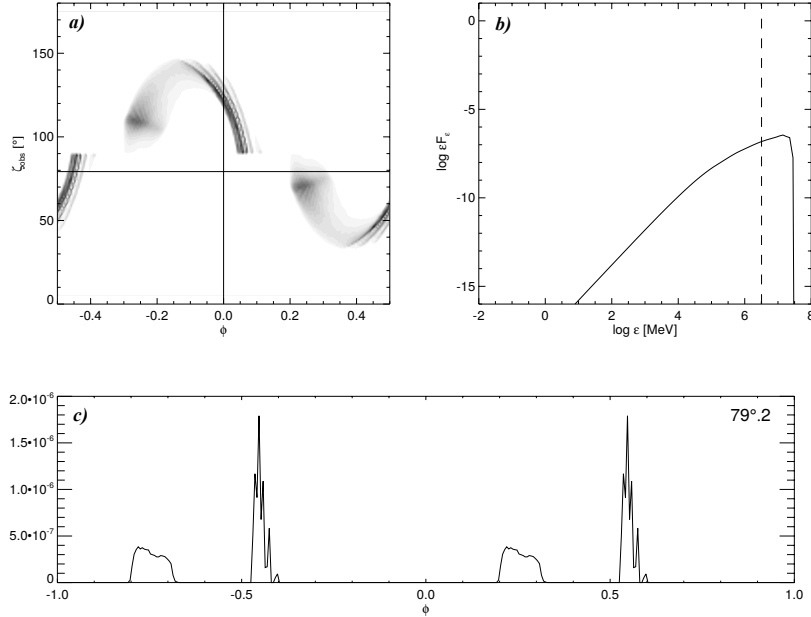}
\vskip -0.5cm
\caption{Panel {\bf a)}: The photon map  shows the distribution of $\sim 3\,$TeV photons due to the ICS between the primary pairs and the optical
radiation from the synchrotron layer for $\alpha = 70\,$deg. The horizontal line is for  $\zeta = 79\,$deg and two periods of rotation are shown for the phase $\phi$ for clarity.
Panel {\bf b)}: The phase-averaged energy spectrum of the ICS component for $\zeta = 79\,$ is shown. The Y-axis scale is in 'model' units.
Panel {\bf c)}: The lightcurve of the ICS at $\sim 3\,$TeV consists of two pulses of different shapes, located at $0.2-0.3$ and $0.5-0.6$ in phase.}
\label{fig2}
\end{figure}

\begin{figure}
\centering
\includegraphics[width=.6\textwidth]{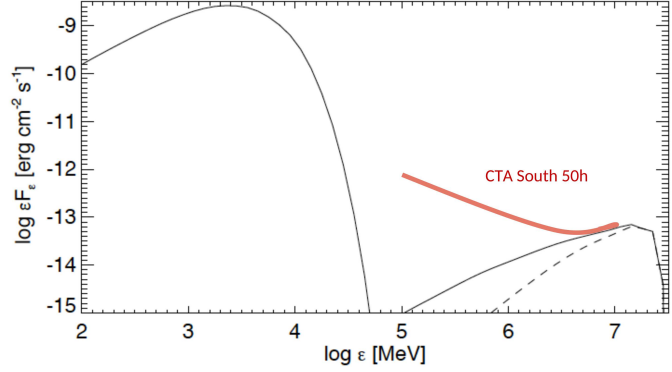}
\vskip -0.5cm
\caption{The phase-averaged SED in HE and VHE range are shown (for $\zeta = 79\,$deg) resulting from the OG model of Vela. The HE part (left-hand side of the panel)
is due to CR of primary pairs in the outer gaps. [Note: see the text for comments on the shape of the CR component]
The VHE component is due to ICS of the SR emission (originating in the SR layer)  on primary pairs. The dashed line component 
corresponds to the SR spectrum extending from
optical to near-IR; the continous line component corresponds to another possibility - with the SR spectrum for the SR spectrum 
assumed  from the optical range up to the far-IR. The red line
indicates the most up-to-date differential sensitivity of the CTA South around the TeV range (\cite{cta}). }
\label{fig3}
\end{figure}

Fig.\ref{fig2} presents the photon map of ICS in the VHE range, as well as the phase averaged spectrum and the lightcurve for the viewing angle
chosen in the previous section and Fig.\ref{fig3}  shows the phase averaged spectra of ICS and CR, with the normalization to match the maximum energy flux of CR 
with the value derived from observations. Note, that an overall shape of the modelled CR component has too little power
at the low-energy part ($\sim 100$ MeV) as well as above  $\sim 10$ GeV. This is a well known fact, and there have been attempts to cure it
by introducing a set of gaps with different external currents and with the time variability at least of the crossing time-scale of 
the light cylinder (\cite{leung}, \cite {takata}). The observed gamma-ray spectrum would then be a superposition of CR components from these gaps.
Other ways out have also been proposed recently (e.g. \cite{akh15}).
For the purpose of this work, the precise shape of the CR component is not essential, however; the key element are ultrarelativistic electrons
in the gap due to an assumed electric field there (see Section 1). \\
Two cases of the input SR spectrum were considered to interact with primary pairs from the gaps via the ICS, both with identical power-law 
model $dN/dE = K(E_{\rm optical} )= 4\, {\rm eV} E^{- \Gamma}$,  with $\Gamma = 1.12$, and the same upper limit of 4 eV, 
but with two different low-energy limits: 1) $\sim 0.1\,$eV in the near-IR (\cite{shib}; see also \cite{danilenko}); 2) $\sim 10^{-3}$eV in the far-IR 
(in principle, one cannot exclude that the 
SR spectrum extends that far, although no observations exist supporting such a statement) .
In the first case, the scatterings proceeds
in the Klein-Nishina regime, and the result is marked with dashed line in Fig.\ref{fig3}.
In the second case, the scattering proceeds in both, the K-N and Thomson, regimes, and the result is marked in Fig.\ref{fig3} with continuous line.

\subsection{A core-like pulse formed in the optical and UV}
\begin{figure}
\centering
\includegraphics[width=.8\textwidth]{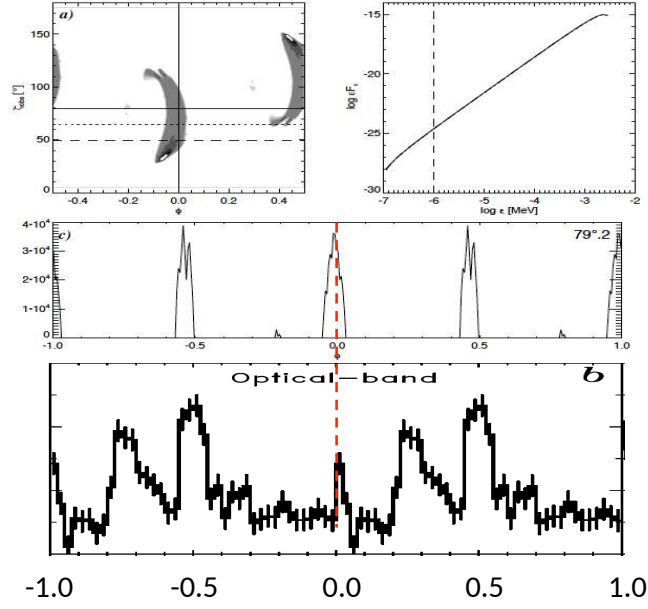}
\vskip -0.5cm
\caption{The photon map  for $\alpha = 70\,$deg shows the distribution of optical photons ($\sim 1\,$eV)
formed due to the ICS between the primary pairs and the optical-to-nearIR
radiation from the synchrotron layer; the continous horizontal line is for  $\zeta = 79\,$deg (top left panel).
The resulting phase-averaged ICS spectrum (note: the Y-axis scale is in 'model' units)
and the ICS lightcurve for $\sim 1\,$eV are included (top right panel and middle panel, respectively).
The bottom panel shows the observed lightcurve of Vela (after \cite{gouiffes}).}
\label{fig4}
\end{figure}

With justification given in Section 1 we assume the presence of inner PC pairs with moderate Lorentz factors ($\sim 100$) outflowing along
the open magnetic-field lines concentrated uniformly around the magnetic axis within a fractional radius $< 0.65$ on the polar cap. The pairs are
followed up to the distance of $0.9\, R_{\rm LC}$  (the light cylinder radius) as they interact via the ICS with photons from the SR layer.
The first case of the SR spectrum mentioned in the previous subsection is used (i.e. with the lower limit at $\sim 0.1\,$eV in the near-IR)
in the computations.
Fig.\ref{fig4} presents the photon map of ICS in the optical range.
Whether the origin
of the so-called P4 peak visible at the phase $\sim 0.0$ is due to the ICS as suggested by the vertical dashed line, remains to be verified by
more detailed numerical calculations.
The second peak in the model, at the phase slightly
below 0.5 is not expected to be detectable as it is overlaid by the peak P2 (of the SR). A similar peak is also formed in the UV range (the spectrum
in Fig.\ref{fig4} extends to the UV and even the Xrays), and it may correspond to P4 visible at the phase 0 in the UV (\cite{uv}.

\subsection{A core-like pulse formed in the hard X-rays}
\begin{figure}
\centering
\includegraphics[width=.7\textwidth]{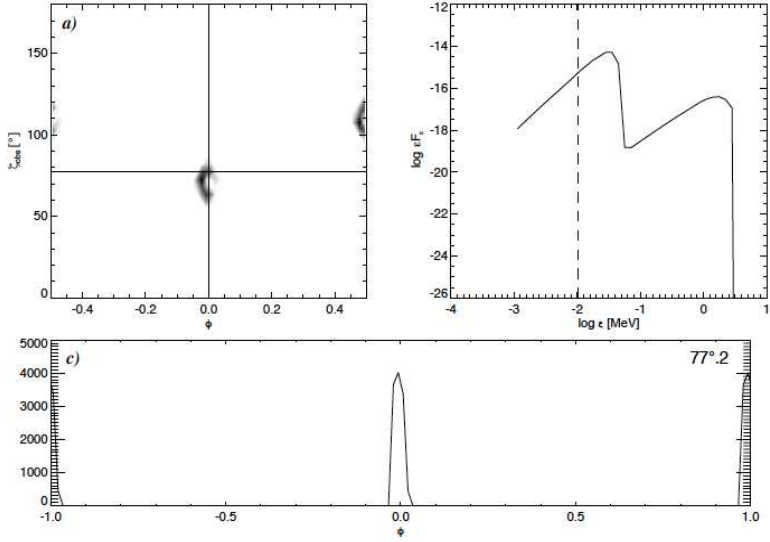}
\vskip -0.5cm
\caption{The photon map  for $\alpha = 70\,$deg shows the distribution of X-ray photons ($\sim 10\,$keV)
formed due to the ICS between the inner PC pairs and the X-rays from the neutron star surface; 
the continous horizontal line is for  $\zeta = 79\,$deg (top left panel).
The resulting phase-averaged ICS spectrum (note: the Y-axis scale is in 'model' units)
and the ICS lightcurve for $\sim 1\,$eV are included (top right panel and bottom panel, respectively).}
\label{fig5}
\end{figure}

The inner PC electron-positron pairs are also capable of scattering thermal photons from the neutron star surface. 
After \cite{manzali} we took a blackbody spectrum with $T_{\rm bb} = 10^6$K and total luminosity $L_{\rm X} = 2 \times 10^{32}$erg/s.
The pairs we followed up to $0.3 R_{\rm LC}$ in the altitude. In the range of interest, between $\sim 1$ keV and $\sim 30$ keV, 
the scatterings  proceed in the magnetic resonance regime (\cite{dermer}), 
forming thus a strong flux of X-ray photons at phase 0.0 (Fig.\ref{fig5}). This may explain the presence of the narrow peak at phase $\sim 0.0$
in the X-ray lightcurves between 0.8 keV and 50 keV (see the compilation of lightcurves in fig.4 of \cite{kuiper}).
 
\subsection{Summary}
The simplified OG model used in this work is the first attempt to account for the presence of specific features in the lightcurves of Vela in optical, UV
and hard-Xrays at the phase where  the radio core peak is located. 
If the origin of the optical and hard-Xray "core" peaks presented above is correct, it will be an addidtional argument (besides the radio core peak)
for an inner gap operating in the central part above the polar cap. The coexistence of inner and outer gaps which occupy different magnetic-field lines 
would have to be taken into account by new models of dissipative magnetospheres.  
 
Moreover, a spectral component
in the VHE gamma-ray domain is found of potential interest to the Cherenkov Telescope Array (CTA), provided
a rather long integrated time of observations (a least 50 hours) is granted (see \cite{cta} for the expected CTA-South sensitivity of point-like sources).
The energy flux of the VHE component is much smaller than derived in the past with 'one-dimensional' models of the OG (\cite{romani},
\cite{aharon}), which also are inconsistent with the HESS upper limits (\cite{hess}). It is, however, much bigger than the flux expected in other models of Vela (e.g. \cite{akh15}, \cite{mochol}).\\
\\
A detailed account of the method and the results presented in this contribution will be given elsewhere (Rudak \& Dyks 2017, in prep.).

\section*{Acknowledgments}
\vskip -0.1cm BR acknowledges discussions with Iwona Mochol and Arache Djannati-Ata\"i.
The authors acknowledge financial support by the National Science Centre grant DEC- 2011/02/A/ST9/00256.

\end{document}